\documentclass[fleqn,10pt]{wlscirep}

\usepackage[utf8]{inputenc}
\usepackage[T1]{fontenc}
\usepackage{babel}

\usepackage{xcolor}

\usepackage{amsmath}
\usepackage{graphicx}
\usepackage{fancyhdr}

\usepackage[flushleft]{threeparttable}

\usepackage{makecell}

\let\vec\mathbf
 
 \usepackage{lineno}

\title{Dictionary of 140k GDB and ZINC derived AMONs}

\author[1,2]{Bing Huang}
\author[1,2,*]{O. Anatole von Lilienfeld}

\affil[1]{
Institute of Physical Chemistry and National Center for Computational Design and Discovery of Novel Materials (MARVEL),
Department of Chemistry, University of Basel, Klingelbergstrasse 80, 4056 Basel, Switzerland
}
\affil[2]{Faculty of Physics, University of Vienna, Vienna, Austria}

\affil[*]{corresponding author(s): O. Anatole von Lilienfeld ( anatole.vonlilienfeld@unibas.ch)}

\begin{abstract}
We present all {\bf A}mons for {\bf G}DB and {\bf Z}inc data-bases using no more than 7 non-hydrogen atoms  (AGZ7)---a calculated organic chemistry building-block dictionary based on the AMON approach [Huang and von Lilienfeld, {\em Nature Chemistry} (2020)]. 
AGZ7 records Cartesian coordinates of compositional and constitutional isomers, as well as properties for $\sim$140k small organic molecules obtained by systematically fragmenting all molecules of Zinc and the majority of GDB17 into smaller entities, saturating with hydrogens, and containing no more than 7 heavy atoms (excluding hydrogen atoms). 
AGZ7 cover the elements \{H, B, C, N, O, F, Si, P, S, Cl, Br, Sn and I\} and includes optimized geometries, total energy and its decomposition, Mulliken atomic charges, dipole moment vectors, quadrupole tensors, electronic spatial extent, eigenvalues of all occupied orbitals, LUMO, gap, isotropic polarizability, harmonic frequencies, reduced masses, force constants, IR intensity, normal coordinates, rotational constants, zero-point energy, internal energy, enthalpy, entropy, free energy, and heat capacity (all at ambient conditions) using B3LYP/cc-pVTZ (pseudopotentials were used for Sn and I) level of theory.
We exemplify the usefulness of this data set with AMON based machine learning models of total potential energy predictions of seven of the most rigid GDB-17 molecules. 
\end{abstract}

\begin{document}

\maketitle
\thispagestyle{empty}

\section*{Background \& Summary}


Chemical compound space (CCS) is vast, primarily due to the combinatorial scaling of the number of possible molecules with number of elements and number of atoms~\cite{anatole-ijqc2013}.
Reliable computational exploration campaigns in CCS  require fast yet accurate computation of molecular properties~\cite{ceder1998predicting,batista2019designmlalchemy}, typically obtained through `on-the-fly' solutions of approximate Schr\"{o}dinger equations, posing a serious bottleneck due to the general scarcity of CPU-time. 
Quantum machine learning (QML) models of quantum properties and applicable throughout CCS, having emerged only over the past decade, can be considered a promising alternative or add-on~\cite{MachineLearningMeetsQuantumPhysics2020book,anatole2020naturereview}. 
 Once properly trained QML models can reach both, high accuracy and efficiency.  The drawback of any QML model is the necessity to acquire training data in sufficient quality and quantity~\cite{QMLessayAnatole}. Once the data has been secured, QML model can solve a wide range of chemical/physical problems within their domain of applicability, ranging from routine quantum chemical calculations (e.g., energy computation, geometry optimization, ab-initio molecular dynamics simulation, etc.) to more application-oriented problems, such as the \emph{de novo} design of molecules or materials exhibiting desired physicochemical properties. As such, the specificity of training data plays a fundamentally important role. However, most \emph{ad-hoc} QML models employ random selection of training data, which represents an important bias since CCS is inhomogeneous. 
It is therefore inevitable that these models will generalize only to a sub-optimal degree, and the training set sizes used are considerably larger than necessary in order to achieve desired accuracy. 
This issue is common to all the existing data-sets in literature, 
e.g.~including the QM$x$ series (QM7b~\cite{qm7b}, QM9~\cite{qm9}, 
QM7-X~\cite{QM7X}, QMspin~\cite{QMspin}, QMrxn~\cite{QMrxn}, QMt~\cite{heinen2020machine}) and (ANI~\cite{ANI1,ANIccx}) series, 
all created based on subsets from the GDB universe~\cite{GDB17,gdb17_exploration}.

Here, we address the issue of bias for rigid and small organic molecules in the gas phase 
by introducing a dictionary of AMONs dubbed ``AGZ7''--- {\bf A}MONs of {\bf G}DB17 and {\bf Z}INC with no more than {\bf 7} non-hydrogen atoms, and covering the relevant local environments present pertinent to organic chemistry, yet maintaining an affordable size. 
AMONs of any query compound include the chemical environments of all the constituting atoms and constitute an improved tailor selected basis within kernel ridge regression (KRR) models of quantum properties which were shown to converge in AMON size and number~\cite{amons}. 
In order to generate a relevant AMON dictionary, we have performed extensive fragmentation campaigns  on two large and diverse exemplary query data-sets GDB-17 ($\sim$137B molecules)~\cite{GDB17,gdb17_exploration} and ZINC ($\sim$817M molecules)~\cite{zinc_db}, covering bio-organic chemistries involving up to 13 chemical elements \{H, B, C, N, O, F, Si, P, S, Cl, Sn, Br, I\}. 
The fragmentation into AMONs has been done in a systematic fashion, accounting for all possible local and unique patterns of atoms in query molecules.
Within AGZ7, all AMONs contains no more than 7 heavy (non-hydrogen) atoms which, as demonstrated previously, enables learning curves to reach mean absolute errors in atomization energies of 1.5 kcal/mol for rigid organic molecules~\cite{amons}.
In order to connect to the predominant body of DFT literature on small organic molecules as well as to the composite quantum chemistry methods developed by Pople, Curtiss and co-workers G$n$-series~\cite{PopleG2,PopleG3,PopleG4}, we have consistently opted for B3LYP/cc-pVTZ as level of theory for all structures and properties.
While the short-comings of common approximations to the exchange-correlation potential in DFT are well known, we note that the fragmentation itself is independent of the electronic structure method, and that it is straightforward to augment and improve upon this level in future studies, e.g.~through the use of multi-level grid combination techniques~\cite{zaspel2018boosting}.
Furthermore, due to their modest size, all AMONs are sufficiently small to remain amenable to more accurate methods, such as CCSD(T)-F12 in a large basis set. 

In this data paper, we report
for approximately 140k AGZ7 molecules (covering the same set of aforementioned 13 elements)
29 physio-chemical properties, including
optimized geometries (based on force-field global minima), fourteen electronic properties (including total energy and its decomposed components, Mulliken atomic charges, dipole moment and its three components, quadruple and its six unique components, isotropic polarizability, HOMO, LUMO, gap energies and eigenvalues of all occupied orbitals, electronic spatial extent) and the other fourteen thermochemical properties (including harmonic frequencies, reduced masses, force constants, IR intensity, normal coordinates, rotational constants, zero-point energy, internal energy, enthalpy, entropy, free energy and heat capacity at ambient conditions, i.e., 298.15 K under 1 atm), and with pseudopotentials being used for Sn and I.
An overview on the GDB and ZINC query and AMON sizes considered is given in Fig.~\ref{fig:size} and Fig.~\ref{fig:statistics}.
Given AGZ7's broad and nearly complete coverage of organic subspace, we believe that this dictionary of relevant AMON-fragments has the potential to become a new standard for next-generation QML models.



This data descriptor is organized into three parts. First, we describe the data source and methods used to generate graph and conformational AMONs. Then we specify  the computational methods used for obtaining geometries and physicochemical properties, as well as some details on the outcome of the computations, followed by a statistical analysis on the AMONs dictionary. Subsequently, we validate AGZ7 by assessing the transferability of local atomic environments to its parent dataset GDB17. Finally, we exemplify the usefuleness of AGZ7 for seven use-cases drawn from GDB-17. 


\section*{Method}

\subsection*{AMON graphs generation}

AMONs generation algorithm~\cite{amons} is used to produce AMON graphs for every query molecule in ZINC or GDB-17, with specific details about the algorithm described in the supplementary material of the original AMONs paper~\cite{amons}. 
Note that for this data-descriptor only SMILES string of query molecules are used as input and accordingly only SMILES strings of smaller representative fragments of queries (aka, AMONs) are generated. 
Note that even if some of the local conformational environment might be missing in the AMONs set of one query molecule, it may still be present in the AMON set of another query molecule. Thus, by combining AMONs of different query molecules or simply using them all as a single training set, the problem inflicted by absence of certain conformational AMONs could be largely overcome.
Furthermore we note that future data-set generation studies will deal more systematically and explicitly with conformational degrees of freedom, e.g.~through inclusion of multiple conformational AMONs corresponding to rotations around rotatable bonds. 

Two sources of query molecules have been used. One corresponds to the GDB17 dataset (a subset of the complete GDB17 dataset~\cite{GDB17}, by courtesy of J.-L. Reymond), consisting of at most 17 heavy atoms and made up of C, H, O, N, F, S, Cl, Br and I, totalling 136'834'437'880 molecules. A more appropriate name for this source dataset would be GBD17s, with `s' indicating ``subset'', but GDB17 is used throughout instead as it will not cause any inconsistency.
The second source is ZINC~\cite{zinc}, a data set consisting of fewer but larger molecules, designed for virtual screening, and representing biologically relevant molecules, totalling 816'777'192
molecules~\cite{zinc_db}. ZINC is also compositionally more diverse than GDB17, with several additional types of atomic environments that are important to organic and bio-chemistry, and including for example >B$-$, >Si<, >P$-$ and >Sn< (see Table~\ref{tab:atomtypes} for more details). 
Charged AMONs and their effect on QML models has not yet been studied in sufficient detail. Therefore, queries (from either GDB17 or ZINC) with non-zero net formal charges have been discarded. 
Any query with partially charged local atoms has been accepted provided that the formal charges appear in pair and opposite in sign, e.g., query involving the functional group -[N+]\#[N-]. Any neutral query molecule with separated formal charges is also discarded as charged AMONs are likely to become necessary.  
By applying the above procedure, we have generated a combined total of 139'643 AMONs, of which 98'218 are unique to ZINC, and 13'048 are unique to GDB17. 28'377 AMONs are shared between these two query datasets.
Resulting filtered query molecules and generated AMONs are all neutral and presumably in their singlet electronic ground state.

\begin{figure} 
\centering
\includegraphics[scale=1]{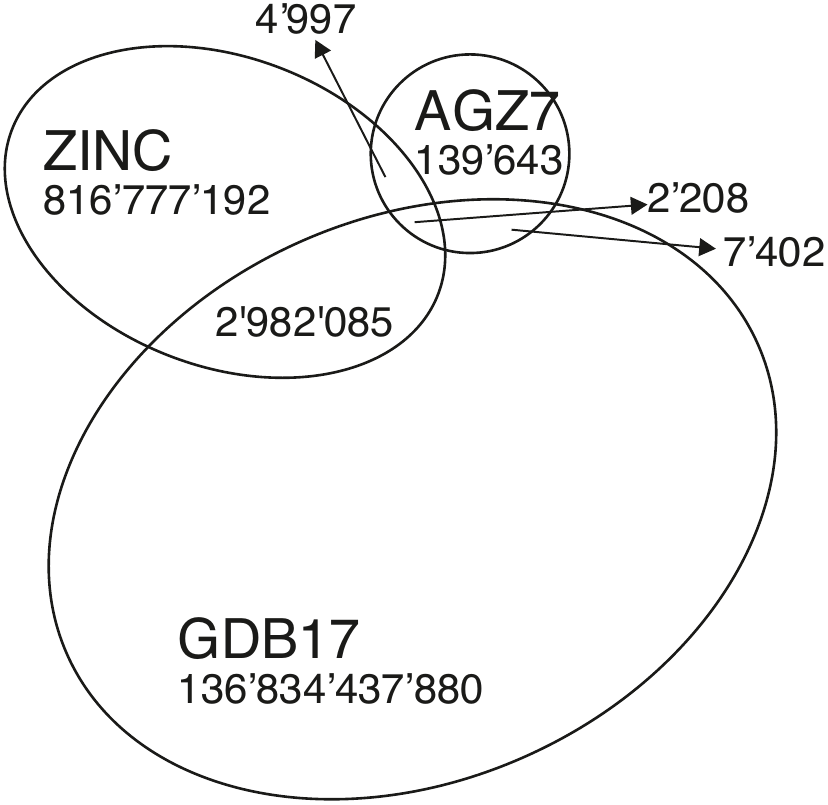}
\caption{Venn diagram for GDB17, ZINC and AGZ7. Inset number shows the size of associated dataset or the number of molecules shared between different datasets. }
\label{fig:venn}
\end{figure}

We note by passing that the shared number of molecules between AGZ7 and ZINC is only a fraction of AGZ7, as is displayed by the Venn diagram in Fig.~\ref{fig:venn}. The same also holds true for GDB17. These observations suggest that the majority of chemistires in GDB17/ZINC cannot be recovered by the small molecules thereof. That is, severe bias is expected for selection of all small molecules (of AMON size, $N_I \leq 7$) from within GDB17/ZINC. To fix this bias by inclusion of random larger molecules, however, is not feasible as the number of necessary large molecules ($N_I > 7$) could easily reach millions and is therefore unaffordable computationally. We will see shortly that AGZ7 is free from such problem.

\subsection*{Search for force field global minima}
As mentioned in the preceding section, we have relied on graphs for the AMON fragmentation, thereby avoiding the conformational complexity. 
In order to build a ``universal'' AMONs set that approximates all conformers on average while still being sufficiently representative,
we have imposed the constraint of only one conformer per AMON graph and we have chosen
the natural option of using the global minima of a given molecular graph.

The identification of global minima from a given molecular graph represents a long-standing challenge in computational chemistry. Many relevant methods have been proposed, e.g.~the minima hopping approach~\cite{minima_hopping}. For our problem at hand, however, we have opted to use a coarse ``global minima'' determined through force field relaxation of generically generated coordinates. 
More specifically, we have employed RDKit~\cite{rdkit}  where both universal force field (UFF)~\cite{uff} and MMFF94 force field~\cite{MMFF94} --- the two methods used in this section --- are efficiently implemented.
The following workflow has been used to identify an estimated lowest conformer for a given graph:
\begin{itemize}
\item[i)] We estimate the number of conformers $n_c$ given a SMILES string by the following rules: A specific rotation number (RN) is firstly assigned to each rotatable bond consisting of heavy atoms, i.e., 3 for a sp$^3$-sp$^3$ bond, 2 for sp$^2$-sp$^2$ and sp$^2$-sp$^3$ type of bonds and 1 for the remaining types of bonds.
Whenever any of the atoms in a bond is within a rigid ring, the corresponding rotation number is determined as $max (\rm{RN}-1, 1)$. Then $n_c$ is calculated as $2\times \prod_i {\rm RN}_i$ with $i$ runs through all bond indices.
\item[ii)] Then $n_c$ conformers are embedded into the molecule using the robust ETKDG~\cite{etkdg} method implemented in RDKit, which uses basic chemical knowledge for bonding and experimental torsion angle preferences, as well as a threshold value of 1.0 for root mean squared distance between 3D coordinates of conformers.
\item[iii)] The resulting conformers are further optimized by MMFF94 and the lowest energy conformer is selected as the approximate global minima.
In the case of MMFF94 force-field parameters being absent for specific atom environments in some molecules, UFF is used instead.
Note that there exists certain high-strain molecules failing step ii) with RDKit, and in this case we have used Openbabel~\cite{obabel} to generate one single conformer and then further optimize its structure by MMFF94 in RDKit.
\end{itemize}
MMFF94 optimized structures are then subsequently used as input of geometry optimization by DFT, as described in the next section.
We find that MMFF94 geometries are reasonable starting points. Geometries optimized by the semi-empirical level of theory PM7 as was implemented in MOPAC2016~\cite{mopac}, however, are not noticeably better for many of the AMONs.
Typically, within DFT an additional set of $\sim$20 ionic steps based on MMFF94 minimas is needed to reach convergence.


\subsection*{Computational method}
B3LYP functional (with the flavour Vosko-Wilk-Nusair functional, type V~\cite{vwn5}, default in MOLPRO~\cite{MOLPRO_brief}) plus D3 dispersion correction~\cite{dft_d3_grimme2010} is used for geometry optimization.
Note that in Gaussian 09 (hereafter G09 for short)~\cite{Gaussian09}, however, another flavour of VWN functional (type III) is used by default.
To be consistent with MOLPRO, an \texttt{iop} entry has to be specified together with \texttt{bv5lyp} in G09 input file, i.e., \texttt{iop(3/76=1000002000, 3/77=0720008000, 3/78=0810010000) bv5lyp)}. 
Targeting accurate geometries, a medium-sized basis set cc-pVTZ is used throughout, except for Sn and I, for which a pseudo-potential version of cc-pVTZ (i.e., cc-pVTZ-PP) is chosen.

To speed up computations, density fitting (df)~\cite{densityfitting_molpro} is used, with df basis cc-pVQZ-JKFIT treating both Coulomb and exchange integrals.
In case cc-pVQZ-JKFIT is not available (e.g., for Sn and I), Def2-QZVPP-JKFIT is chosen instead. With this setting, the difference in computed energies is within sub-kJ/mol compared to energies without density fitting.

The MOLPRO 2018 package~\cite{MOLPRO_brief} is employed for almost all geometry optimizations.
Default setting on geometry convergence thresholds are used (max total energy change: $10^{-6}$ Hartree, max force: 3$\times 10^{-4}$ Hartree/Bohr, max displacement: 3$\times 10^{-4}$ Bohr).
When geometry optimization is hard to converge (e.g.~for very flexible molecules),
and considering that calculation of the Hessian is too expensive to conduct for better determination of the next move of ions, we have chosen to run MOLPRO and G09 in an alternate way: First, 12 steps of geometry optimization by df-B3LYP is carried out in MOLPRO,
based on the energy gradient of the last configuration, 
the Berny optimization algorithm implemented in G09 is used 
to determine the next most probable configuration.
Then, MOLPRO carries on with energy and force computation, 
afterwards G09 for ionic move, $\cdots$, and so on and so forth, until convergence is reached.

Subsequently, frequency calculations have been carried out for all AMONs by G09 with all details on method identical to the above except that no density fitting is used (not implemented for B3LYP in G09).
Switching to G09 is desirable as the computation of analytical Hessian by hybrid DFT (B3LYP) is available. 
Calculations of all other properties aforementioned follow once a frequency calculation is done.
All thermochemical properties were calculated at standard ambient conditions, i.e., 298.15 K and 1 atm.

Out of the overall 139'643 AMONs, 157 were found to be dissociated after geometry optimization and thus have been excluded for further frequency calculation.
The remaining ones include 5'684 AMONs which underwent a change in their molecular graph.
In the \texttt{readme.txt} file of the data respositary (see the next section), all AMONs that have experienced rearrangement of connectivity have been recorded and are specified using their indices in the database.





\section*{Data Records}
The resulting AMONs dataset is named ``AGZ7'' for convenience of reference in later sections and future work, where the letters in the name indicate respectively the words AMONs, GDB17 and ZINC and the number 7 alludes to the maximal number of heavy atoms involved.
Accordingly, ``AG7'' and ``AZ7'' are used to indicate AMONs set of GDB17 and ZINC, respectively. 

Optimized molecular geometries and their associated properties are presented in plain text-based extended xyz format (see the File format subsection for details),
publicly available at GitHub (\texttt{https://github.com/binghuang2018/agz7}). 


\subsection*{Properties}
All properties have been calculated at the DFT/B3LYP/cc-pVTZ level
of theory on the relaxed geoemtry (optimized at the same level), using pseudo-potential options for 
Sn and I. Overall 28 properties have been obtained for $\sim$140k AMON molecules, as listed in Table~\ref{tab:extxyz}.

\subsection*{File format}
Extended XYZ file format is used for storing all data of a molecule. Associated file name is of the format \texttt{dataset\_index.xyz}, where \texttt{dataset} could be one of \{\texttt{ag7}, \texttt{az7}, \texttt{agz7}\}.

A typical XYZ file consists of 3 parts: header (the first line, i.e., number of atoms), comment (the second line) and atomic coordinates (third to last line, one atom per line). We extend this format by assigning as comment a series of scalar properties universal to all molecules once a single point DFT calculation is done and appending another series of properties unique to a G09 frequency calculation. For details, see Table~\ref{tab:extxyz}.

\section*{Technical Validation}


\subsection*{Statistics}
\begin{figure} 
\centering
\includegraphics[scale=0.1]{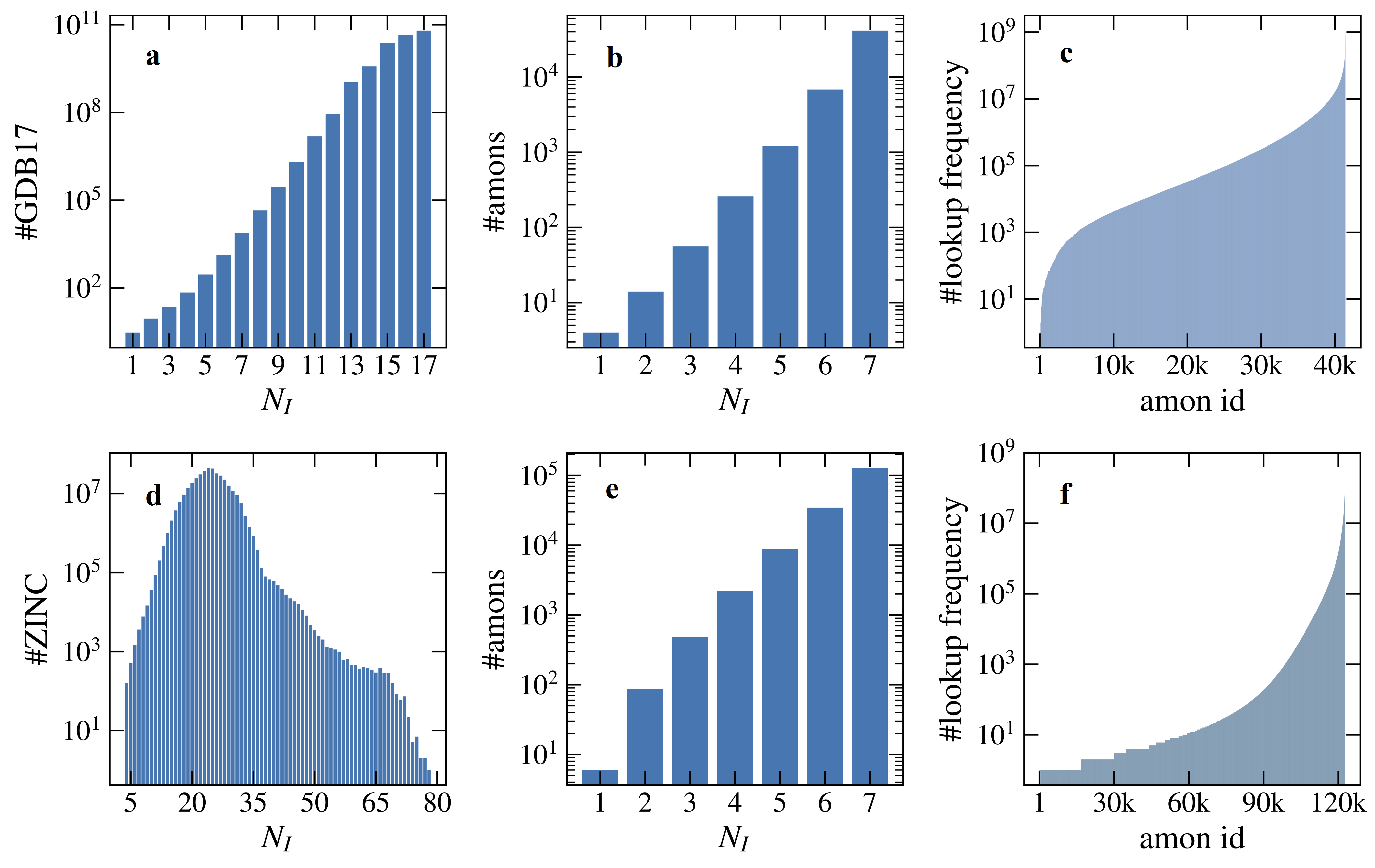}
\caption{Statistics on GDB17, ZINC and respective AMONs. {\bf a}. Distribution of GDB17 molecules over different values of $N_I$. $N_I$: number of non-hydrogen atoms in a molecule, \#GDB17: number of GDB17 molecules; {\bf b}. The same type of plot as {\bf a} but for AG7 molecules; {\bf c}. The frequency of AG7 molecules to be looked up by query GDB17 molecules, sorted in ascending order. {\bf d}-{\bf f} display information similar to {\bf a}-{\bf c} respectively, except that the query dataset is ZINC. }
\label{fig:statistics}
\end{figure}

As a first step, we have analyzed the statistics of both ZINC and GDB17, as well as the look-up frequency of their AMONs.
As displayed in Fig.~\ref{fig:statistics}a, the number of GDB17 molecules roughly increases exponentially w.r.t. the number of heavy atoms ($N_I$) when $N_I \le 14$. The non-exponential increase above $N_I=15$ is due to the introduction of additional graph and functional group selections that Reynmond's group implemented gradually at 15, 16 and 17 nodes to reduce the explosion in numbers,
as was explained in details in reference \cite{GDB17}.
For ZINC, the number of molecules peaks at $N_I=24$ and then decreases almost exponentially w.r.t. $N_I$. 
Compared to GDB17, the distribution of ZINC molecules (Fig.~\ref{fig:statistics}d) is conspicuous and suggests severe bias, probably arising as a consequence of manual preparation of the dataset,  instead of automatized exhaustion of possible combinations of local atomic environments based on a consistent set of rules. 

Not surprisingly, these differences are smoothened in the corresponding AMONs statistics as AMONs themselves correspond to exhaustive enumeration of all constituting local environments of larger query molecules,
no matter how many query molecules there are or how they are distributed over different $N_I$'s.
More specifically, the number of AMONs increases exponentially w.r.t. $N_I$ for both AG7 (GDB17 AMONs) and AZ7 (ZINC AMONs), differing only in magnitude (see Fig.~\ref{fig:statistics}b and d). 

One last noticeable difference between GDB17 and ZINC is the look-up frequency of respective AMONs.
Apparently, AG7 AMONs are generally more frequently looked up due to i) fewer atom types are involved and ii) queries basically correspond to all possible enumerations of these atom types.
On the contrary, ZINC look-ups from AZ7 correspond to sparse combinations of more atom types (see Table~\ref{tab:atomtypes}), favoring certain atom types, e.g., >C<, >C= and >N-.
Therefore, striking difference exists in the distribution of look-up frequency of AMONs of the two datasets:
the distribution is much flatter for GDB17 than for ZINC, with almost one half of ZINC AMONs looked up no more than 10 times. As such, these distributions provide powerful indicators on the homogeneity of the CCS under study.

\begin{figure}  
\centering
\includegraphics[scale=0.35]{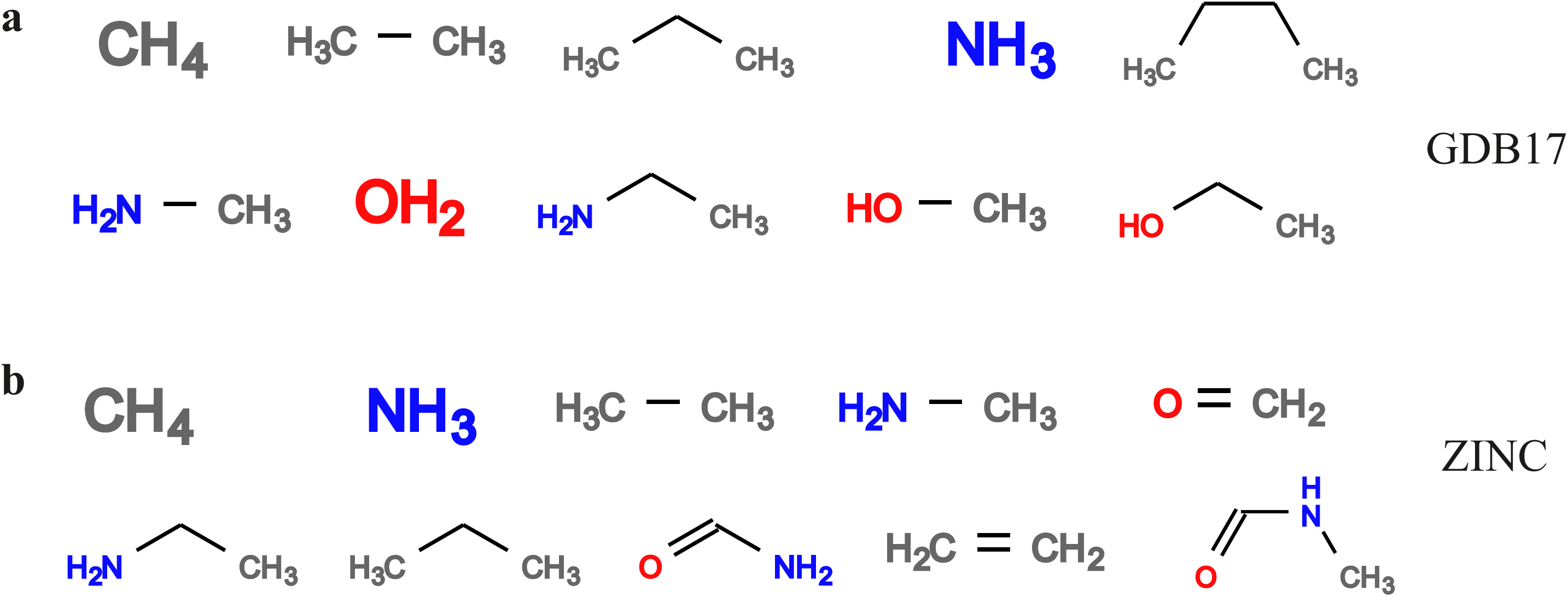}
\caption{The ten most frequent AMONs of GDB17 ({\bf a}) and ZINC ({\bf b}) datasets, sorted in descending order by look-up frequency (from left to right, up to down). 
}
\label{fig:AMONs10}
\end{figure}

The preference of certain atomic environments could be further confirmed by comparing the most frequently looked-up AMONs, as displayed in Fig.~\ref{fig:AMONs10}.
For both AZ7 and AG7, AMONs with SMILES strings \texttt{C}, \texttt{CC}, \texttt{N} and \texttt{NCC} are the most fundamental ones.
Dominating AMONs in AZ7 include \texttt{O=CNC} and its smaller constituting fragments, i.e., \texttt{O=C}, \texttt{N}, \texttt{C}, \texttt{NC} and \texttt{O=CN}. 
For AG7, the 10 most frequently looked-up AMONs are all saturated.
On the contrary, bio-active species favor more unsaturated building blocks, e.g., \texttt{C=C} and \texttt{C=O}, as relevant bond (e.g., a neighboring C-N single bond in the local environment >C(=O)N<) is amenable to transformation in bio-active species in pertinent bio-chemical reactions.

\subsection*{Coverage analysis} 
\begin{figure}  
\centering
\includegraphics[scale=0.7]{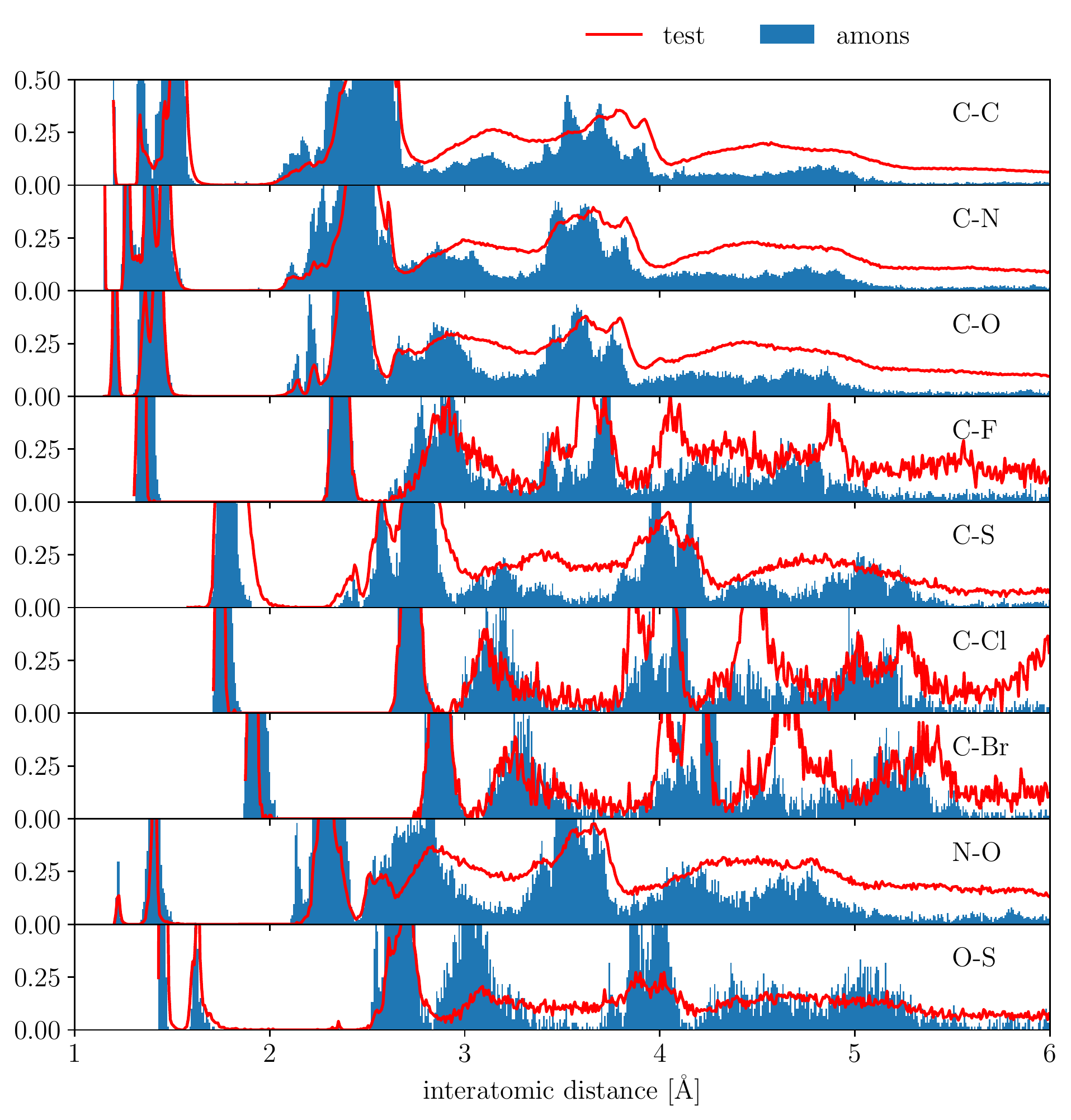}
\caption{Interatomic distance distribution (normalized to 1) of AG7 (as AMONs, blue) and GDB17 (as test molecules, red) for 9 atomic pairs as indicated within each subplot. 
}
\label{fig:2body_all}
\end{figure}

\begin{figure}  
\centering
\includegraphics[scale=0.7]{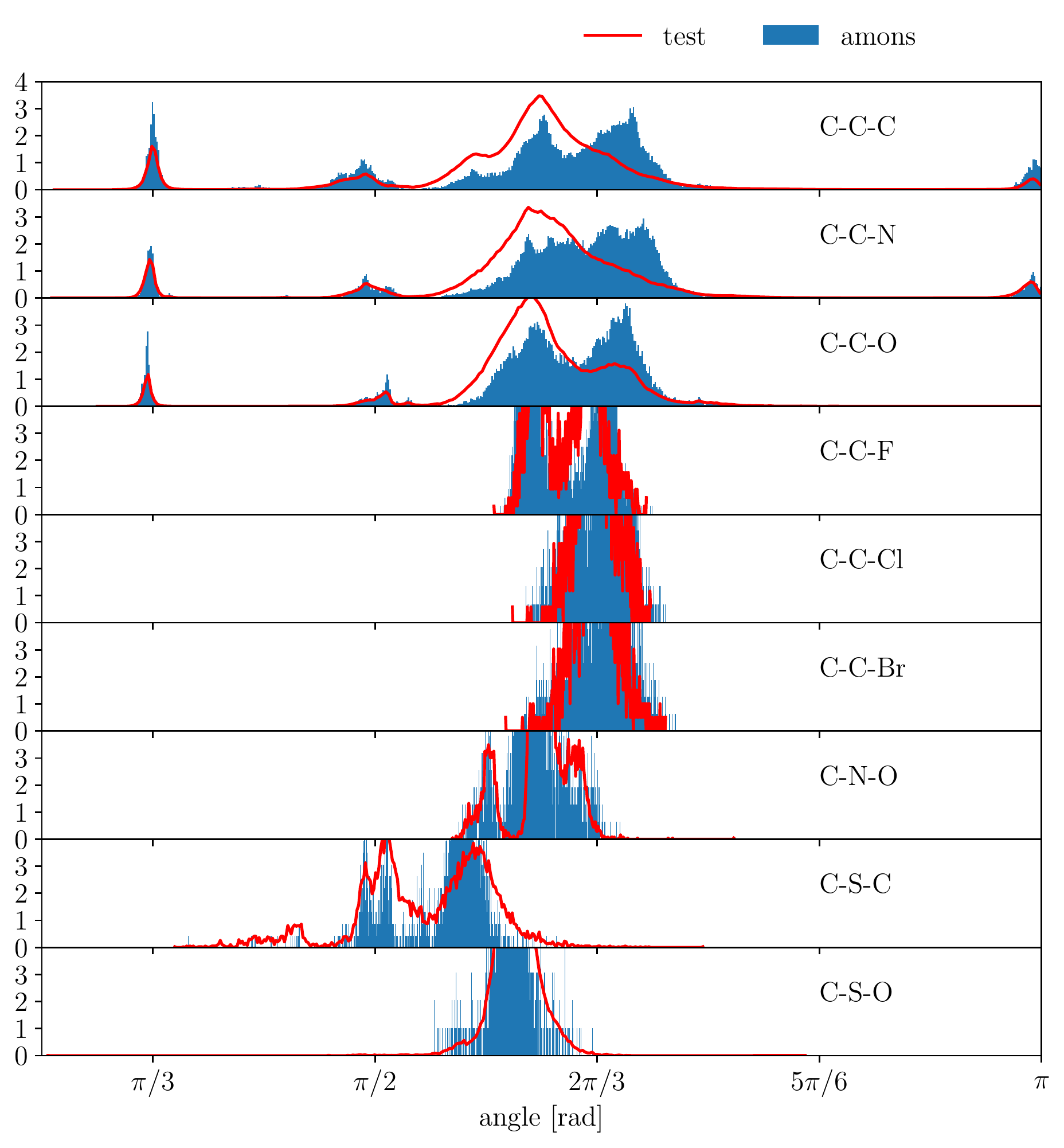}
\caption{Angular distribution (normalized to 1) of AG7 (as AMONs, blue) and GDB17 (as test molecules, red) for 9 angles as indicated within each subplot. Note that each angle is spanned by two adjacent covalent bonds, sharing one atom in common.
}
\label{fig:3body_all}
\end{figure}

Here, we roughly assess the generalizability of the training set (AG7) by checking
if the local atomic environments of AG7 cover exactly that of an associated test set (consisting of $\sim$100k molecules with $N_I>=10$, drawn randomly from GBD17).
This has be done by analyzing the distribution of 2-body and 3-body terms in both datasets.
The distribution of interatomic distances (2-body term) are displayed in Fig.~\ref{fig:2body_all} for a total of 9 atomic pairs for both AMONs (AG7) and test GDB17 molecules.
As expected, the domain of chemical space spanned by the test molecules fully overlap with that of the AMONs,
though the sampling frequency of different region in chemical space is quite different.
This perfect match of coverage unambiguously conveys the message that unbiased sampling is roughly achieved, rendering possible high accuracy prediction of properties of test molecules with minimal cost.
Fig.~\ref{fig:3body_all} shows comparison of distributions involving the angles (3-body term) for the same two datasets. Similar observations can be made as for the two-body case. 

\subsection*{Use-case demonstration}
The AMON data-set introduced above can be used in many ways. 
One obvious application is to treat AGZ7 as a dictionary from which training-instances are selected ``on-the-fly'' for training a specific target query molecule. 
More specifically, given the graph of a new query molecule,
one generates its corresponding graph AMONs by the AMON generation program~\cite{aqmlcode} and then look up the properties of these AMONs from AGZ7.
One then builds a QML model (e.g., kernel ridge regression plus some robust molecular representaion such as aSLATM~\cite{amons} or FCHL~\cite{FCHL}) on the resulting small set of data.
Once trained, the properties of the target molecule can be predicted. 

\begin{figure}  
\centering
\includegraphics[scale=0.8]{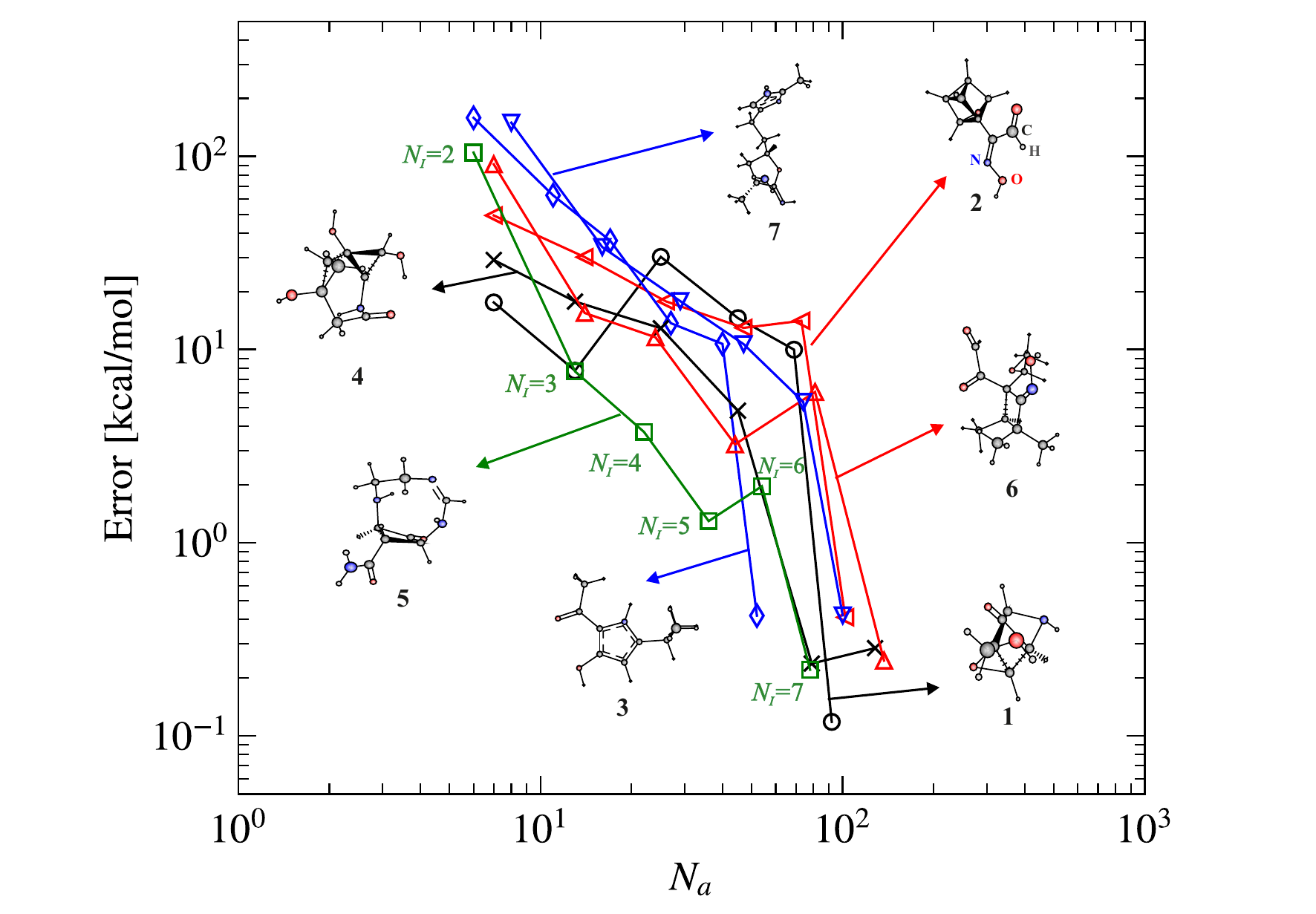}
\caption{Learning curves shown as B3LYP total energy prediction error (ordinate, in kcal/mol) plotted against the number of training AMONs (abscissa, $N_a$) for each of the seven rigid GDB17 molecules (labeled by bold integers ranging from 1 to 7) containing respectively 10-16 heavy atoms. A KRR-aSLATM QML model~\cite{amons} was used for training and test. $N_I=m$ indicates the subset of amons containing no more than $m$ heavy atoms.}
\label{fig:lc}
\end{figure}

To illustrate this use case, we have drawn 7 of the most rigid molecules (i.e., the molecules that  have the smallest difference between the number of complete conformer amons and graph amons) from GDB17 dataset,
and plotted their AMON based learning curves of the predicted total energy error in Fig.~\ref{fig:lc}.
As can be seen from the figure, very low absolute prediction errors are rapidly achieved as corresponding AMONs grow in size and number. 


\bibliography{reference}

\begin{thebibliography}{10}
\urlstyle{rm}
\expandafter\ifx\csname url\endcsname\relax
  \def\url#1{\texttt{#1}}\fi
\expandafter\ifx\csname urlprefix\endcsname\relax\def\urlprefix{URL }\fi
\expandafter\ifx\csname doiprefix\endcsname\relax\def\doiprefix{DOI: }\fi
\providecommand{\bibinfo}[2]{#2}
\providecommand{\eprint}[2][]{\url{#2}}

\bibitem{anatole-ijqc2013}
\bibinfo{author}{von Lilienfeld, O.~A.}
\newblock \bibinfo{journal}{\bibinfo{title}{First principles view on chemical
  compound space: Gaining rigorous atomistic control of molecular properties}}.
\newblock {\emph{\JournalTitle{Int. J. Quantum Chem.}}}
  \textbf{\bibinfo{volume}{113}}, \bibinfo{pages}{1676--1689}
  (\bibinfo{year}{2013}).

\bibitem{ceder1998predicting}
\bibinfo{author}{Ceder, G.}
\newblock \bibinfo{journal}{\bibinfo{title}{Predicting properties from
  scratch}}.
\newblock {\emph{\JournalTitle{Science}}} \textbf{\bibinfo{volume}{280}},
  \bibinfo{pages}{1099--1100} (\bibinfo{year}{1998}).

\bibitem{batista2019designmlalchemy}
\bibinfo{author}{Freeze, J.~G.}, \bibinfo{author}{Kelly, H.~R.} \&
  \bibinfo{author}{Batista, V.~S.}
\newblock \bibinfo{journal}{\bibinfo{title}{Search for catalysts by inverse
  design: Artificial intelligence, mountain climbers, and alchemists}}.
\newblock {\emph{\JournalTitle{Chemical reviews}}}  (\bibinfo{year}{2019}).

\bibitem{MachineLearningMeetsQuantumPhysics2020book}
\bibinfo{author}{Sch{\"u}tt, K.} \emph{et~al.}
\newblock \emph{\bibinfo{title}{Machine Learning Meets Quantum Physics}}.
\newblock Lecture Notes in Physics (\bibinfo{publisher}{Springer International
  Publishing}, \bibinfo{year}{2020}).

\bibitem{anatole2020naturereview}
\bibinfo{author}{von Lilienfeld, O.~A.}, \bibinfo{author}{Tkatchenko, A.} \&
  \bibinfo{author}{M{\"u}ller, K.-R.}
\newblock \bibinfo{journal}{\bibinfo{title}{Exploring chemical compound space
  with quantum-based machine learning}}.
\newblock {\emph{\JournalTitle{arXiv preprint arXiv:1911.10084}}}
  (\bibinfo{year}{2019}).

\bibitem{QMLessayAnatole}
\bibinfo{author}{von Lilienfeld, O.~A.}
\newblock \bibinfo{journal}{\bibinfo{title}{Quantum machine learning in
  chemical compound space}}.
\newblock {\emph{\JournalTitle{Angew. Chem. Int. Ed.}}}
  \textbf{\bibinfo{volume}{57}}, \bibinfo{pages}{4164--4169}
  (\bibinfo{year}{2018}).

\bibitem{qm7b}
\bibinfo{author}{Montavon, G.} \emph{et~al.}
\newblock \bibinfo{journal}{\bibinfo{title}{Machine learning of molecular
  electronic properties in chemical compound space}}.
\newblock {\emph{\JournalTitle{New J. Phys.}}} \textbf{\bibinfo{volume}{15}},
  \bibinfo{pages}{095003} (\bibinfo{year}{2013}).

\bibitem{qm9}
\bibinfo{author}{Ramakrishnan, R.}, \bibinfo{author}{Dral, P.},
  \bibinfo{author}{Rupp, M.} \& \bibinfo{author}{von Lilienfeld, O.~A.}
\newblock \bibinfo{journal}{\bibinfo{title}{Quantum chemistry structures and
  properties of 134 kilo molecules}}.
\newblock {\emph{\JournalTitle{Sci. Data}}} \textbf{\bibinfo{volume}{1}},
  \bibinfo{pages}{140022} (\bibinfo{year}{2014}).

\bibitem{QM7X}
\bibinfo{author}{Hoja, J.} \emph{et~al.}
\newblock \bibinfo{journal}{\bibinfo{title}{{QM7-X}: A comprehensive dataset of
  quantum-mechanical properties spanning the chemical space of small organic
  molecules}}.
\newblock {\emph{\JournalTitle{arXiv preprint arXiv:2006.15139}}}
  (\bibinfo{year}{2020}).

\bibitem{QMspin}
\bibinfo{author}{Schwilk, M.}, \bibinfo{author}{Tahchieva, D.~N.} \&
  \bibinfo{author}{von Lilienfeld, O.~A.}
\newblock \bibinfo{journal}{\bibinfo{title}{Large yet bounded: Spin gap ranges
  in carbenes}}.
\newblock {\emph{\JournalTitle{arXiv preprint arXiv:2004.10600}}}
  (\bibinfo{year}{2020}).

\bibitem{QMrxn}
\bibinfo{author}{von Rudorff, G.~F.}, \bibinfo{author}{Heinen, S.~N.},
  \bibinfo{author}{Bragato, M.} \& \bibinfo{author}{von Lilienfeld, O.~A.}
\newblock \bibinfo{journal}{\bibinfo{title}{Thousands of reactants and
  transition states for competing e2 and sn2 reactions}}.
\newblock {\emph{\JournalTitle{arXiv preprint arXiv:2006.00504}}}
  (\bibinfo{year}{2020}).

\bibitem{heinen2020machine}
\bibinfo{author}{Heinen, S.}, \bibinfo{author}{Schwilk, M.},
  \bibinfo{author}{von Rudorff, G.~F.} \& \bibinfo{author}{von Lilienfeld,
  O.~A.}
\newblock \bibinfo{journal}{\bibinfo{title}{Machine learning the computational
  cost of quantum chemistry}}.
\newblock {\emph{\JournalTitle{Machine Learning: Science and Technology}}}
  \textbf{\bibinfo{volume}{1}}, \bibinfo{pages}{025002} (\bibinfo{year}{2020}).

\bibitem{ANI1}
\bibinfo{author}{Smith, J.~S.}, \bibinfo{author}{Isayev, O.} \&
  \bibinfo{author}{Roitberg, A.~E.}
\newblock \bibinfo{journal}{\bibinfo{title}{Ani-1, a data set of 20 million
  calculated off-equilibrium conformations for organic molecules}}.
\newblock {\emph{\JournalTitle{Sci. Data}}} \textbf{\bibinfo{volume}{4}},
  \bibinfo{pages}{170193} (\bibinfo{year}{2017}).

\bibitem{ANIccx}
\bibinfo{author}{Smith, J.~S.} \emph{et~al.}
\newblock \bibinfo{journal}{\bibinfo{title}{The ani-1ccx and ani-1x data sets,
  coupled-cluster and density functional theory properties for molecules}}.
\newblock {\emph{\JournalTitle{Scientific Data}}} \textbf{\bibinfo{volume}{7}},
  \bibinfo{pages}{1--10} (\bibinfo{year}{2020}).

\bibitem{GDB17}
\bibinfo{author}{Ruddigkeit, L.}, \bibinfo{author}{van Deursen, R.},
  \bibinfo{author}{Blum, L.~C.} \& \bibinfo{author}{Reymond, J.-L.}
\newblock \bibinfo{journal}{\bibinfo{title}{Enumeration of 166 billion organic
  small molecules in the chemical universe database gdb-17}}.
\newblock {\emph{\JournalTitle{J. Chem. Inf. Model.}}}
  \textbf{\bibinfo{volume}{52}}, \bibinfo{pages}{2864--2875}
  (\bibinfo{year}{2012}).

\bibitem{gdb17_exploration}
\bibinfo{author}{Fink, T.}, \bibinfo{author}{Bruggesser, H.} \&
  \bibinfo{author}{Reymond, J.-L.}
\newblock \bibinfo{journal}{\bibinfo{title}{Virtual exploration of the
  small-molecule chemical universe below 160 daltons}}.
\newblock {\emph{\JournalTitle{Angew. Chem. Int. Ed.}}}
  \textbf{\bibinfo{volume}{44}}, \bibinfo{pages}{1504--1508}
  (\bibinfo{year}{2005}).

\bibitem{amons}
\bibinfo{author}{Huang, B.} \& \bibinfo{author}{von Lilienfeld, O.~A.}
\newblock \bibinfo{journal}{\bibinfo{title}{The {``DNA''} of chemistry:
  {Scalable} quantum machine learning with ``amons''}}.
\newblock {\emph{\JournalTitle{arXiv preprint arXiv:1707.04146}}}
  (\bibinfo{year}{2017}).

\bibitem{zinc_db}
\bibinfo{note}{Http://zinc15.docking.org/tranches/home/. Last retrieval date:
  Dec 22, 2018.}

\bibitem{PopleG2}
\bibinfo{author}{Curtiss, L.~A.}, \bibinfo{author}{Raghavachari, K.},
  \bibinfo{author}{Trucks, G.~W.} \& \bibinfo{author}{Pople, J.~A.}
\newblock \bibinfo{journal}{\bibinfo{title}{Gaussian-2 theory for molecular
  energies of first-and second-row compounds}}.
\newblock {\emph{\JournalTitle{J. Chem. Phys.}}} \textbf{\bibinfo{volume}{94}},
  \bibinfo{pages}{7221--7230} (\bibinfo{year}{1991}).

\bibitem{PopleG3}
\bibinfo{author}{Curtiss, L.~A.}, \bibinfo{author}{Redfern, P.~C.},
  \bibinfo{author}{Raghavachari, K.}, \bibinfo{author}{Rassolov, V.} \&
  \bibinfo{author}{Pople, J.~A.}
\newblock \bibinfo{journal}{\bibinfo{title}{Gaussian-3 theory using reduced
  mo/ller-plesset order}}.
\newblock {\emph{\JournalTitle{The Journal of chemical physics}}}
  \textbf{\bibinfo{volume}{110}}, \bibinfo{pages}{4703--4709}
  (\bibinfo{year}{1999}).

\bibitem{PopleG4}
\bibinfo{author}{Curtiss, L.~A.}, \bibinfo{author}{Redfern, P.~C.} \&
  \bibinfo{author}{Raghavachari, K.}
\newblock \bibinfo{journal}{\bibinfo{title}{Gaussian-4 theory}}.
\newblock {\emph{\JournalTitle{The Journal of chemical physics}}}
  \textbf{\bibinfo{volume}{126}}, \bibinfo{pages}{084108}
  (\bibinfo{year}{2007}).

\bibitem{zaspel2018boosting}
\bibinfo{author}{Zaspel, P.}, \bibinfo{author}{Huang, B.},
  \bibinfo{author}{Harbrecht, H.} \& \bibinfo{author}{von Lilienfeld, O.~A.}
\newblock \bibinfo{journal}{\bibinfo{title}{Boosting quantum machine learning
  models with multi-level combination technique: Pople diagrams revisited}}.
\newblock {\emph{\JournalTitle{Journal of chemical theory and computation}}}
  (\bibinfo{year}{2018}).

\bibitem{zinc}
\bibinfo{author}{Irwin, J.~J.} \& \bibinfo{author}{Shoichet, B.~K.}
\newblock \bibinfo{journal}{\bibinfo{title}{Zinc- a free database of
  commercially available compounds for virtual screening}}.
\newblock {\emph{\JournalTitle{J. Chem. Inf. Model.}}}
  \textbf{\bibinfo{volume}{45}}, \bibinfo{pages}{177--182}
  (\bibinfo{year}{2005}).

\bibitem{minima_hopping}
\bibinfo{author}{Goedecker, S.}
\newblock \bibinfo{journal}{\bibinfo{title}{Minima hopping: An efficient search
  method for the global minimum of the potential energy surface of complex
  molecular systems}}.
\newblock {\emph{\JournalTitle{J. Chem. Phys.}}}
  \textbf{\bibinfo{volume}{120}}, \bibinfo{pages}{9911--9917}
  (\bibinfo{year}{2004}).

\bibitem{rdkit}
\bibinfo{note}{RDKit: Open-source cheminformatics; http://www.rdkit.org}.

\bibitem{uff}
\bibinfo{author}{Rappe, A.~K.}, \bibinfo{author}{Casewit, C.~J.},
  \bibinfo{author}{Colwell, K.~S.}, \bibinfo{author}{III, W. A.~G.} \&
  \bibinfo{author}{Skiff, W.~M.}
\newblock \bibinfo{journal}{\bibinfo{title}{Uff, a full periodic table force
  field for molecular mechanics and molecular dynamics simulations}}.
\newblock {\emph{\JournalTitle{J. Am. Chem. Soc.}}}
  \textbf{\bibinfo{volume}{114}}, \bibinfo{pages}{10024--10035}
  (\bibinfo{year}{1992}).

\bibitem{MMFF94}
\bibinfo{author}{Halgren, T.~A.}
\newblock \bibinfo{journal}{\bibinfo{title}{Merck molecular force field. i.
  basis, form, scope, parameterization, and performance of mmff94}}.
\newblock {\emph{\JournalTitle{J. Comp. Chem.}}} \textbf{\bibinfo{volume}{17}},
  \bibinfo{pages}{490--519} (\bibinfo{year}{1996}).

\bibitem{etkdg}
\bibinfo{author}{Riniker, S.} \& \bibinfo{author}{Landrum, G.~A.}
\newblock \bibinfo{journal}{\bibinfo{title}{Better informed distance geometry:
  using what we know to improve conformation generation}}.
\newblock {\emph{\JournalTitle{J. Chem. Inf. Model.}}}
  \textbf{\bibinfo{volume}{55}}, \bibinfo{pages}{2562--2574}
  (\bibinfo{year}{2015}).

\bibitem{obabel}
\bibinfo{author}{O'Boyle, N.~M.} \emph{et~al.}
\newblock \bibinfo{journal}{\bibinfo{title}{Open babel: An open chemical
  toolbox}}.
\newblock {\emph{\JournalTitle{J. Cheminform.}}} \textbf{\bibinfo{volume}{3}},
  \bibinfo{pages}{1--14} (\bibinfo{year}{2011}).

\bibitem{mopac}
\bibinfo{note}{MOPAC2016, James J. P. Stewart, Stewart Computational Chemistry,
  Colorado Springs, CO, USA, HTTP://OpenMOPAC.net (2016).}

\bibitem{vwn5}
\bibinfo{author}{Vosko, S.~H.}, \bibinfo{author}{Wilk, L.} \&
  \bibinfo{author}{Nusair, M.}
\newblock \bibinfo{journal}{\bibinfo{title}{Accurate spin-dependent electron
  liquid correlation energies for local spin density calculations: a critical
  analysis}}.
\newblock {\emph{\JournalTitle{Can. J. Phys.}}} \textbf{\bibinfo{volume}{58}},
  \bibinfo{pages}{1200--1211} (\bibinfo{year}{1980}).

\bibitem{MOLPRO_brief}
\bibinfo{author}{Werner, H.-J.} \emph{et~al.}
\newblock \bibinfo{title}{Molpro, version 2018.1, a package of ab initio
  programs} (\bibinfo{year}{2015}).

\bibitem{dft_d3_grimme2010}
\bibinfo{author}{Grimme, S.}, \bibinfo{author}{Antony, J.},
  \bibinfo{author}{Ehrlich, S.} \& \bibinfo{author}{Krieg, H.}
\newblock \bibinfo{journal}{\bibinfo{title}{A consistent and accurate ab initio
  parametrization of density functional dispersion correction (dft-d) for the
  94 elements h-pu}}.
\newblock {\emph{\JournalTitle{J. Chem. Phys.}}}
  \textbf{\bibinfo{volume}{132}}, \bibinfo{pages}{154104}
  (\bibinfo{year}{2010}).

\bibitem{Gaussian09}
\bibinfo{author}{Frisch, M.~J.} \emph{et~al.}
\newblock \bibinfo{title}{Gaussian~09 {R}evision {D}.01}.
\newblock \bibinfo{note}{Gaussian Inc. Wallingford CT 2009}.

\bibitem{densityfitting_molpro}
\bibinfo{author}{Werner, H.-J.}, \bibinfo{author}{Manby, F.~R.} \&
  \bibinfo{author}{Knowles, P.~J.}
\newblock \bibinfo{journal}{\bibinfo{title}{Fast linear scaling second-order
  m{\o}ller-plesset perturbation theory (mp2) using local and density fitting
  approximations}}.
\newblock {\emph{\JournalTitle{J. Chem. Phys.}}}
  \textbf{\bibinfo{volume}{118}}, \bibinfo{pages}{8149--8160}
  (\bibinfo{year}{2003}).

\bibitem{aqmlcode}
\bibinfo{author}{Huang, B.} \& \bibinfo{author}{von Lilienfeld, O.~A.}
\newblock \bibinfo{title}{Aqml: Amons-based quantum machine learning code for
  quantum chemistry, https://github.com/binghuang2018/aqml}
  (\bibinfo{year}{2020}).

\bibitem{FCHL}
\bibinfo{author}{Faber, F.~A.}, \bibinfo{author}{Christensen, A.~S.},
  \bibinfo{author}{Huang, B.} \& \bibinfo{author}{von Lilienfeld, O.~A.}
\newblock \bibinfo{journal}{\bibinfo{title}{Alchemical and structural
  distribution based representation for universal quantum machine learning}}.
\newblock {\emph{\JournalTitle{J. Chem. Phys.}}}
  \textbf{\bibinfo{volume}{148}}, \bibinfo{pages}{241717}
  (\bibinfo{year}{2018}).

\end{thebibliography}

\section*{Acknowledgements}
Both authors greatly acknowledge J.~L.~Reymond for sharing the GDB-17 SMILES strings.
This work was mainly being funded by the Swiss National Science foundation through 407540\_167186 NFP 75 Big Data. O.A.v.L. also acknowledges funding from the Swiss National Science foundation (200021\_175747, NCCR MARVEL) and from the European Research Council (ERC-CoG grant QML).
Some calculations were performed at sciCORE (http://scicore.unibas.ch/) scientific computing core facility at University of Basel.

\section*{Author contributions}
Both authors conceived of the presented idea. 
B.H. implemented the methods and carried out all calculations. 
Both authors discussed the results and wrote the paper.

\section*{Competing interests}
The authors declare no competing financial interests.

\newpage

\begin{table}[]			
\caption {Atom types in GDB17 and ZINC AMONs. Atom or atom types that do not appears in GDB17 are indicated by the superscript $^a$. Atom type is encoded as \{formal charge\}\_\{BO$_1$\}\{BO$_1$\}$\cdots$\{BO$_n$\}, where ``BO'' is bond order and $n$ is the total number of neigbhors of the atom considered. Note that BO's are sorted in descending order. For instance, atom type ``0\_2211'' of sulfer is present in the molecule CS(=O)(=O)C. 
}
\label{tab:atomtypes}
\begin{center}
\begin{tabular}{llllllllll}	 \hline
Atom & Type & & & & & & & \\  \hline			
H	&	0\_1	&		&		&		&		&		&		&	\\
B$^a$	&	0\_111	&	0\_21	&	-1\_211	&	-1\_1111	&		&		&		&	\\
C	&	0\_1111	&	0\_211	&	0\_22	&	0\_31	&	-1\_4$^a$	&		&		&	\\
N	&	0\_111	&	0\_21	&	0\_3	&	1\_22	&	1\_211	&	1\_31	&	1\_1111	&	-1\_2\\
O	&	0\_11	&	0\_2	&	-1\_1	&		&		&		&		&	\\
F	&	0\_1	&		&		&		&		&		&		&	\\
Si$^a$	&	0\_1111	&	0\_211	&		&		&		&		&		&	\\
P$^a$	&	0\_111	&	0\_21	&	0\_3	&	0\_2111	&	0\_221	&		&		&	\\
S	&	0\_11	&	0\_2	&	0\_211	&	0\_222	&	0\_2211	&		&		&	\\
Cl	&	0\_1	&		&		&		&		&		&		&	\\
Ge$^a$	&	0\_1111	&		&		&		&		&		&		&	\\
Sn$^a$	&	0\_1111	&	0\_211	&		&		&		&		&		&	\\
Br	&	0\_1	&		&		&		&		&		&		&	\\
I	&	0\_1	&		&		&		&		&		&		&	\\ \hline
\end{tabular}											\end{center}			
\end{table}																

\begin{table}
\begin{center}
\begin{threeparttable}
\caption{Extended XYZ file format for molecular structure and properties} \label{tab:extxyz}
\begin{tabular}{lllll}	 \hline
Line number & Cotent & Property & Label & Note\\  \hline			
1           & number of atoms & $N$ & & \\  
2           & identity & & \texttt{id} & e.g., AG7\_00001 \\ 
            & SMILES & & \texttt{smi} & B3LYP geometry \\ 
            & number of heavy atoms & $N_{I}$ & \texttt{ni} &  \\ 
            & number of electrons & $n_e$ & \texttt{nelec} &  \\ 
            & total energy & $E$ & \texttt{et} & in Hartree \\ 
            & nucleus-electron interaction energy & $E_{ne}$ & \texttt{ene} & in Hartree \\ 
            & nuclear repulsion energy & $E_{nn}$ & \texttt{enn} & in Hartree \\ 
            & D3 dispersion energy & $E_{\rm D3}$ & \texttt{ed3} & in Hartree \\ 
            & kinetic energy & $E_{\rm k}$ & \texttt{ek} & in Hartree \\ 
 & dipole moment & $\mu$ or $|\vec{\mu}|$  & \texttt{mu} & in Debye \\
 & $\vec{\mu}$ components & $\mu_x$  & \texttt{mux} & in Debye \\
 & & $\mu_y$  & \texttt{muy} & in Debye \\
 & & $\mu_z$  & \texttt{muz} & in Debye \\
 & quadrupole moment & $Q$ \\ 
 & $Q$ components & $Q_{xx}$  & \texttt{qxx} & in Debye-Ang \\
 & & $Q_{yy}$  & \texttt{qyy} & in Debye-Ang \\
 & & $Q_{zz}$  & \texttt{qzz} & in Debye-Ang \\
 & & $Q_{xy}$  & \texttt{qyy} & in Debye-Ang \\
 & & $Q_{xz}$  & \texttt{qzz} & in Debye-Ang \\
 & & $Q_{yz}$  & \texttt{qzz} & in Debye-Ang \\
 & HOMO energy & $\varepsilon_{\rm{HOMO}}$ & \texttt{homo}  & in Hartree \\
 & LUMO energy & $\varepsilon_{\rm{LUMO}}$ & \texttt{lumo}  & in Hartree \\ 
 & HOMO-LUMO gap & $\Delta\varepsilon$ & \texttt{gap} & in Hartree \\
 & electronic spatial extent & $\langle R^2 \rangle$ & \texttt{r2} & in Bohr$^2$ \\
3, $\dots$, $N$+2 & element \& coordinates & & & in Angstrom, column 1-4 \\ 
   & Mulliken atomic charge & $q_A$ & \texttt{qa} & in $e$, column 5 \\
   & atomic mass & $m_A$ & \texttt{ma} & in Dalton, column 6 \\
 $N$+3 & occupied orbital eigenvalues & $\varepsilon$ & \texttt{eocc} & in Hartree, $n_e$/2 entries \\ 
\hline  
    $N$+4 & harmonic frequencies & $\omega$ & \texttt{omega} & in cm$^{-1}$, let \#entries=$n_{\omega}$ \\ 
    $N$+5 & reduced masses & $m_{r}$ & \texttt{mr} & Dalton, $n_{\omega}$ entries\\ 
    $N$+6 & force constants & $k$ & \texttt{fc} & in mDyne/\AA, $n_{\omega}$ entries \\ 
    $N$+7 & IR intensity & $I$ & \texttt{ir} & in km/mol, $n_{\omega}$ entries \\ 
    $N$+8 & normal coordinates & $\mathbf{R}_{norm}$ & \texttt{rn} & within $[0,1]$, $n_{\omega}\times 3N$ entries\\ 
    $N$+9 & highest $\omega$ & $\omega_0$ & \texttt{oemga0} & in cm$^{-1}$ \\ 
      & rotational constant & A & \texttt{a} & in GHz \\  
      & rotational constant & B & \texttt{b} & in GHz \\  
      & rotational constant & C & \texttt{c} & in GHz \\  
      & isotropic polarizability & $\alpha$ & \texttt{alpha} & in Bohr$^3$ \\  
    & zero point energy & ZPE & \texttt{zpe} & in kcal/mol\\
    & internal energy at 0 K & $U_0$ & \texttt{u0} & in Hartree\\ 
    & internal energy at 298.15 K & $U$ & \texttt{u} & in Hartree\\ 
    & enthalpy at 298.15 K & $H$ & \texttt{h} & in Hartree \\
    & free energy at 298.15 K & $G$ & \texttt{g} & in Hartree\\ 
    & entropy at 298.15 K & $S$ & \texttt{s} & in cal/mol/K\\
    & heat capacity at 298.15 K & $C_{\rm{v}}$ & \texttt{cv} & in cal/mol/K\\ \hline 
\end{tabular}
\end{threeparttable}
\end{center}
\end{table}

\end{document}